\documentclass[a4paper,11pt]{article}                        
\usepackage[affil-it]{authblk}
\usepackage{graphics} 
\usepackage{relsize}
\usepackage{subfigure}
\usepackage{graphicx,color}
\usepackage{amsmath} 
\usepackage{amssymb}  
\usepackage{amsmath,amssymb,amsfonts,theorem}
\usepackage{enumerate}
\usepackage{hyperref}
\usepackage{epstopdf}
\graphicspath{ {images/} }
\usepackage{xcolor}
\usepackage{pgfplots} 
\pgfplotsset{compat=newest} 
\pgfplotsset{plot coordinates/math parser=false} 
\newlength\figureheight 
\newlength\figurewidth

\def\undertilde#1{\mathord{\vtop{\ialign{##\crcr
$\hfil\displaystyle{#1}\hfil$\crcr\noalign{\kern1.5pt\nointerlineskip}
$\hfil\tilde{}\hfil$\crcr\noalign{\kern1.5pt}}}}}

{\theorembodyfont{\normalfont} 

\def\be{\begin{equation}}
\def\ee{\end{equation}}
\def\ba{\begin{array}}
\def\ea{\end{array}}
\def\eqa{\begin{eqnarray}}
\def\eqe{\end{eqnarray}}

\begin{document}
\title{{\LARGE \bf{Modeling of power distribution systems with solar generation: A case study }}}
 
\author{Kees Loeff, Matin Jafarian, Jacquelien M.A. Scherpen}
 
\affil{Engineering and Technology Institute, University of Groningen, The Netherlands}
\date{}
\maketitle
\section{Introduction}
\label{int}
The recent growth of renewable energy sources, i.e. wind and solar sources, in electrical power systems is tremendous. Generation from renewable energy sources is mainly added to the power network in a decentralized fashion, and the availability of the generation depends on weather conditions. One of the most common renewable generation techniques is photovoltaic (solar) generation. 

Regarding the challenges of integrating renewables in power grid, there has been an increasing interest to study and analyze the behavior and performance of power systems with renewable generations e.g. \cite{matincdc2016}. The proposed methods and algorithms require to be tested by means of simulating appropriate case studies. The purpose of this document is to present a case study of distribution power systems composed of a single-phase distribution bus, solar (PV) inverters and daily load profiles based on real and actual data.

The distribution system is modeled as a modified IEEE 37 bus for which the modified data for single-phase analysis are provided. Moreover, a simplified model for PV inverter is used to obtain desired active and reactive output power given the weather condition, e.g. temperature and solar radiation. The modeling and data are based on \cite{kees}.\\[1mm]
This document is organized as follows. Section \ref{sec:case} presents the data for a modified single-phase distribution network based on IEEE 37 benchmark model. Section \ref{sec:load} gives the load profile for active and reactive power based on the daily use of an average Dutch household. The model and data for the simplified PV inverter is given in Section \ref{sec:photo}. The note is concluded in Section \ref{sec:con}.
\section{A modified IEEE 37 bus}\label{sec:case}
We take an IEEE 37 bus \cite{ieee37} to model the distribution system. The IEEE 37 bus is a three-phase, unbalanced medium voltage ($4.8$ kV) network. However, many analytical problems assume a three-phase balanced network that allows to equivalently consider a single-phase network in the analysis. Here, we briefly present the steps to obtain a single-phase network based on IEEE 37 and provide the data for the single-phase bus. We modify the line data (series impedances and suceptances) using the symmetrical components method \cite{kersting2012distribution} in order to obtain a three-phase balanced network in two steps. First, the lines are assumed to be transposed \cite{kersting2012distribution}. Next, the sequence impedance matrix is calculated (see \cite{kersting2012distribution} for detailed explanations). Section \ref{sec:I37} presents the bus configuration (Figure \ref{fig:I37}) and the modified line parameters for the single-phase analysis of this distribution bus. 
\subsection{Modified IEEE 37 bus: Phase series impedance and shunt susceptance matrices }\label{sec:I37}
\begin{figure}[h]
\centering
\includegraphics [width=0.56\linewidth]{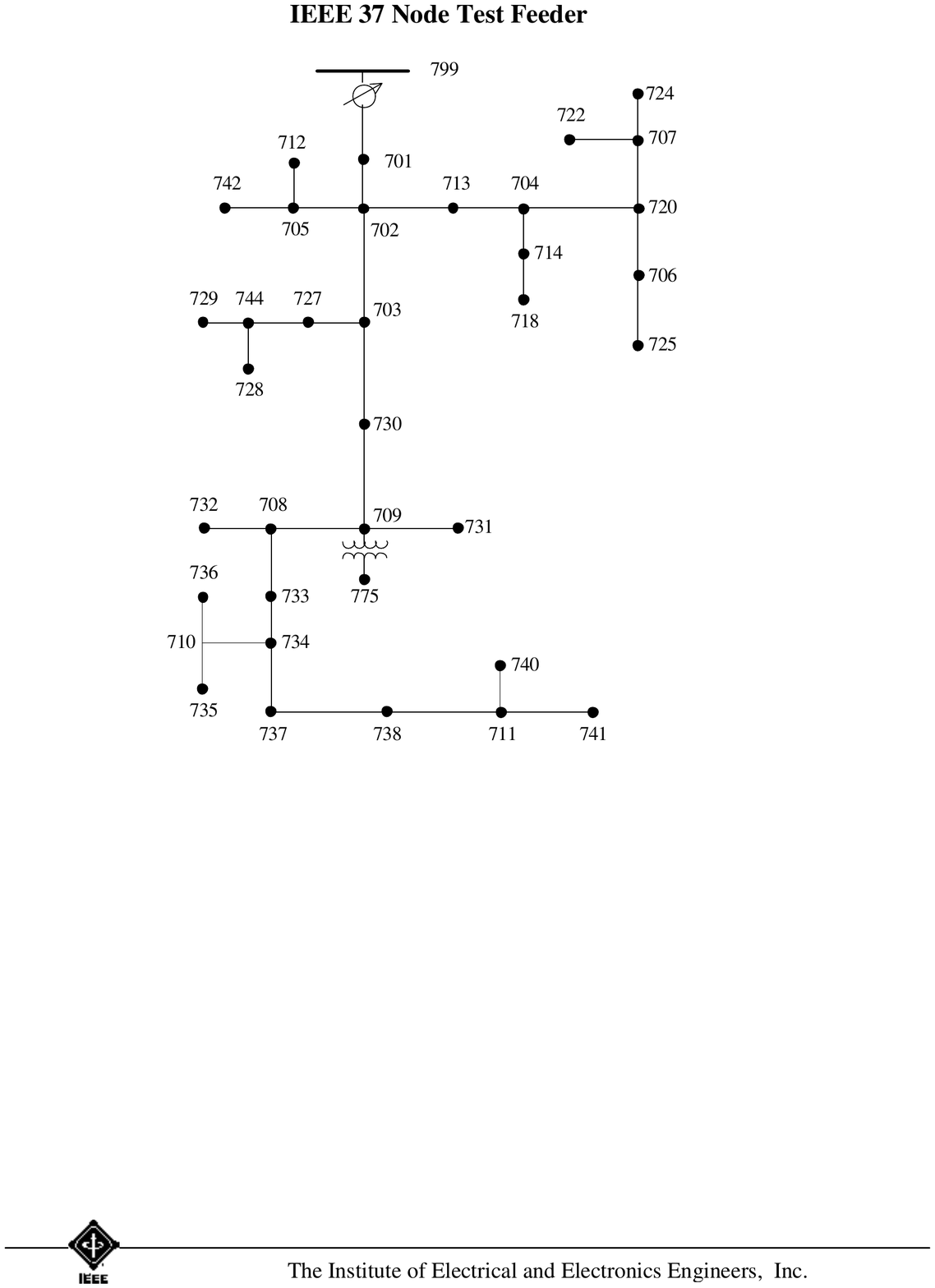}
\caption{IEEE 37 bus (node) test feeder.}
\label{fig:I37}
\end{figure}

Here we present the data for a modified three-phase balanced network based on IEEE 37-bus (see \cite{ieee37} for the original data) based on \cite{kees}. Figure \ref{fig:I37} shows the configuration of the bus. We assume that the transformer between busbars 709 and 775 is a one-to-one transformer.
\subsection{Line configurations}
\textbf{Configuration 721}\\
 \textit{Phase series impedance, $Z = R + jX$ in [Ohms/mile]}
\begin{equation*}
\begin{bmatrix}
0.2926 + j 0.1973 & 0.0673 -j 0.0368 & 0.0337 - j 0.0417\\
0.0673 -j 0.0368 & 0.2646 + j 0.1900 & 0.0673 - j 0.0368\\
0.0337 - j 0.0417 & 0.0673 - j 0.0368 & 0.2926 + j 0.1973
 \end{bmatrix}
\end{equation*}
\textit{Shunt susceptance, $B$ in [$\mu$Siemens/mile]}\\
\begin{equation*}
\begin{bmatrix}
159.7919 & 0.0000 & 0.0000\\
 0.0000 & 159.7919 & 0.0000\\
0.0000 & 0.0000 & 159.7919
\end{bmatrix}
\end{equation*}
\textbf{Configuration 722}\\
\textit{Phase series impedance, $Z = R + jX$ in [Ohms/mile]}
\begin{equation*}
\begin{bmatrix}
0.4751 + j 0.2973 & 0.1629 -j 0.0326 & 0.1234 - j 0.0607\\
0.1629 -j 0.0326 & 0.4488 + j 0.2678 & 0.1629 - j 0.0326\\
0.1234 - j 0.0607 & 0.1629 - j 0.0326 & 0.4751 + j 0.2973
\end{bmatrix}
\end{equation*}
\textit{Shunt susceptance, $B$ in [$\mu$Siemens/mile]}
\begin{equation*}
\begin{bmatrix}
127.8306 & 0.0000 & 0.0000\\
0.0000 & 127.8306 & 0.0000\\
0.0000 & 0.0000 & 127.8306
\end{bmatrix}
\end{equation*}
\textbf{Configuration 723}\\
\textit{Phase series impedance, $Z = R + jX$ in [Ohms/mile]}
\begin{equation*}
\begin{bmatrix}
1.2936 + j 0.6713 & 0.4871 +j 0.2111 & 0.4585 + j 0.1521\\
0.4871 +j 0.2111 & 1.3022 + j 0.6326 & 0.4871 - j 0.2111\\
0.4585 + j 0.1521 & 0.4871 - j 0.2111 & 1.2936 + j 0.6713
\end{bmatrix}
\end{equation*}
\textit{Shunt susceptance, $B$ in [$\mu$Siemens/mile]}
\begin{equation*}
\begin{bmatrix}
74.8405 & 0.0000 & 0.0000\\
0.0000 & 74.8405 & 0.0000\\
0.0000 & 0.0000 & 74.8405
\end{bmatrix}
\end{equation*}
\textbf{Configuration 724}\\
\textit{Phase series impedance, $Z = R + jX$ in [Ohms/mile]}
\begin{equation*}
\begin{bmatrix}
2.0952 + j 0.7758 & 0.5204 + j 0.2738 & 0.4926 + j 0.2123\\
0.5204 +j 0.2738 & 2.1068 + j 0.7398 & 0.5204 + j 0.2738\\
0.4926 + j 0.2123 & 0.5204 + j 0.2738 & 2.0952 + j 0.7758
\end{bmatrix}
\end{equation*}
\textit{Shunt susceptance, $B$ in [$\mu$Siemens/mile]}
\begin{equation*}
\begin{bmatrix}
60.2483 & 0.0000 & 0.0000\\
0.0000 & 60.2483 & 0.0000\\
0.0000 & 0.0000 & 60.2483
\end{bmatrix}
\end{equation*}
\pagebreak
\section{Loads}\label{sec:load}
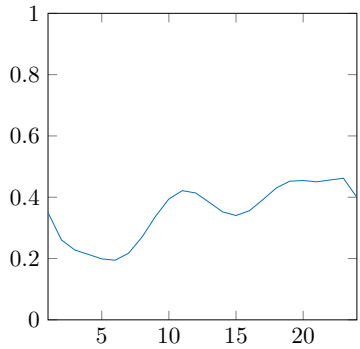
\begin{figure}[h]
\begin{center}
\definecolor{mycolor1}{rgb}{0.00000,0.44700,0.74100} 
\begin{tikzpicture}[scale=0.8]
\begin{axis}[%
width=2in,
height=2in,
at={(0.758in,0.481in)},
scale only axis,
xmin=1,
xmax=24,
ymin=0,
ymax=1,
axis background/.style={fill=white}
]
\addplot [color=mycolor1,solid,forget plot]
  table[row sep=crcr]{%
1	0.34989618091757\\
2	0.260764569295833\\
3	0.227625082616227\\
4	0.213599514761665\\
5	0.19896437265222\\
6	0.194544413700119\\
7	0.217345573817537\\
8	0.270123824760642\\
9	0.337320390737621\\
10	0.394348686434529\\
11	0.421468931064124\\
12	0.41391015696956\\
13	0.383576670059692\\
14	0.352290228383214\\
15	0.34035647198798\\
16	0.356193541553001\\
17	0.392314336823079\\
18	0.430206352582157\\
19	0.452302245608344\\
20	0.454580878738046\\
21	0.450286098631457\\
22	0.456305363133077\\
23	0.462021867722932\\
24	0.399654244876938\\
};
\end{axis}
\end{tikzpicture}\end{center}
\caption{Daily load profile for normal electricity use for a household in month June with an annual electricity demand of 3400 kWh.}
\label{fig:load}
\end{figure}

Loads are modeled as hourly constant $P$ and $Q$ (active and reactive power). We adopt the profile of the loads during a day in the month June from \cite{veldman2010modelling} as shown in Figure \ref{fig:load}. This profile represents the active power consumption of an average Dutch household. Since IEEE 37 bus is a medium voltage bus, we multiply the load profile of one household by 20 in order to have the magnitude of the net active power load for each busbar in the same order as the IEEE 37 bus. Furthermore, we take the reactive power load as $50\%$ of the active power load, similar to the IEEE 37 data.
\section{PV inverter: model and data}\label{sec:photo}
\begin{figure}[h]
\centering
\includegraphics [width=0.7\linewidth]{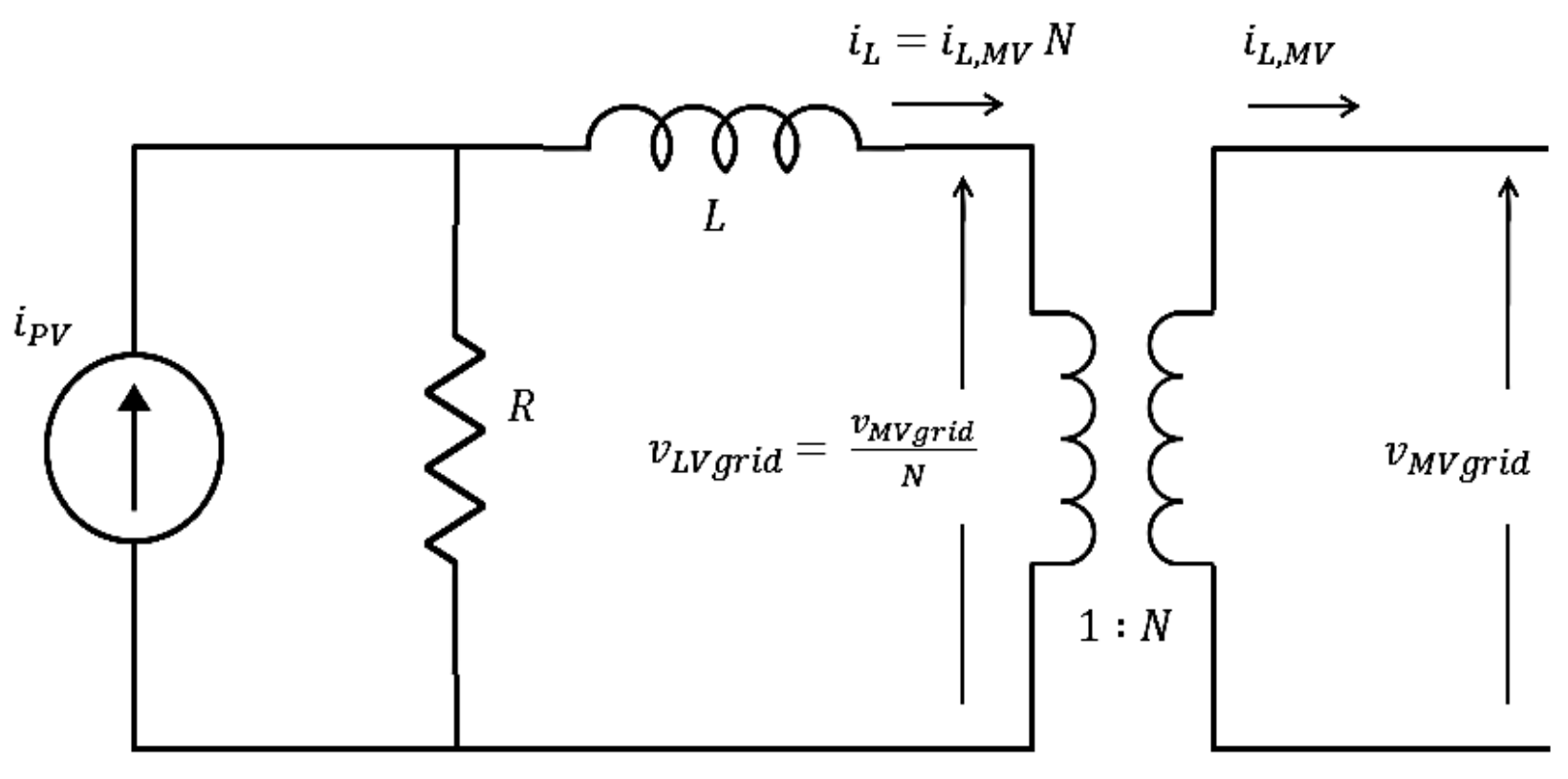}
\caption{The simplified model of the PV inverter.}
\label{fig:pv1}
\end{figure}

\begin{figure}[h]
\centering
\includegraphics [width=0.6\linewidth]{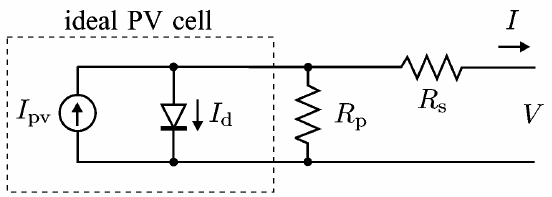}
\caption{PV-cell equivalent circuit \cite{villalva2009comprehensive}.}
\label{fig:pv2}
\end{figure}

\begin{figure}[h]
\centering
\includegraphics[scale=.7]{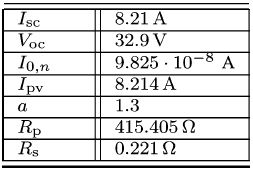}
\caption{Parameters of KC200GT Solar Array at 298 [$K$], AM1.5 (air mass), 1000 [$W/m^2$] \cite{kc200gt}.}
\label{fig:para}
\end{figure}

\begin{figure}[h]
\centering
\includegraphics[scale=.36]{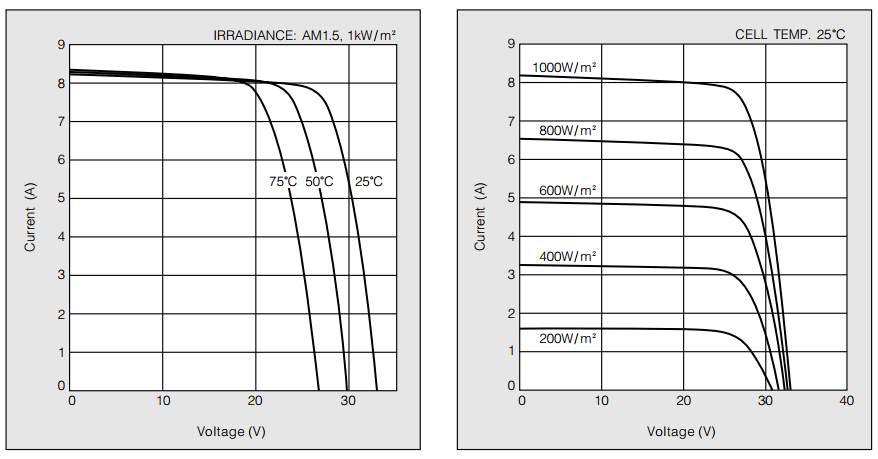}
\caption{I-V curve at constant temperature and different solar irradiation, I-V curve at constant solar irradiation and different temperatures \cite{kc200gt}}
\label{fig:I-Vcurve}
\end{figure}

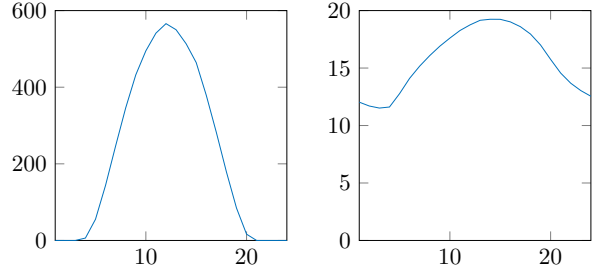
\begin{figure}
\begin{center}
\definecolor{mycolor1}{rgb}{0.00000,0.44700,0.74100}%
\begin{tikzpicture}[scale=0.8]
\begin{axis}[%
width=1.5in,
height=1.5in,
at={(0.758in,0.481in)},
scale only axis,
xmin=1,
xmax=24,
ymin=0,
ymax=600,
axis background/.style={fill=white}
]
\addplot [color=mycolor1,solid,forget plot]
  table[row sep=crcr]{%
1	0\\
2	0\\
3	0\\
4	6.2962962962963\\
5	55.3086419753086\\
6	143.981481481481\\
7	246.851851851852\\
8	346.172839506173\\
9	432.253086419753\\
10	494.938271604938\\
11	541.111111111111\\
12	565.956790123457\\
13	550.061728395062\\
14	513.364197530864\\
15	464.29012345679\\
16	378.888888888889\\
17	281.944444444444\\
18	178.91975308642\\
19	84.6296296296296\\
20	16.358024691358\\
21	0\\
22	0\\
23	0\\
24	0\\
};
\end{axis}
\end{tikzpicture}%
\hspace{3mm}
\definecolor{mycolor1}{rgb}{0.00000,0.44700,0.74100}%
\begin{tikzpicture}[scale=0.8]
\begin{axis}[%
width=1.5in,
height=1.5in,
at={(0.758in,0.481in)},
scale only axis,
xmin=1,
xmax=24,
ymin=0,
ymax=20,
axis background/.style={fill=white}
]
\addplot [color=mycolor1,solid,forget plot]
  table[row sep=crcr]{%
1	12.0322222222222\\
2	11.6966666666667\\
3	11.5188888888889\\
4	11.61\\
5	12.77\\
6	14.1044444444444\\
7	15.1677777777778\\
8	16.0844444444444\\
9	16.9022222222222\\
10	17.6077777777778\\
11	18.2644444444444\\
12	18.7555555555556\\
13	19.1544444444444\\
14	19.2444444444444\\
15	19.24\\
16	19.0177777777778\\
17	18.5988888888889\\
18	17.96\\
19	17.0066666666667\\
20	15.7533333333333\\
21	14.5622222222222\\
22	13.6788888888889\\
23	13.0377777777778\\
24	12.5455555555556\\
};
\end{axis}
\end{tikzpicture}%
\end{center}
\caption{Solar irradiation (left) and temperature (right) for an average day in June based on the data of 2013, 14, 15.}
\label{fig:ts}
\end{figure}

We model each photovoltaic panel with inverter as a simplified circuit shown in Figure \ref{fig:pv1} where the magnitude of current $i_{pv}$ depends on the environmental parameters, e.g. temperature and radiation, determined by a photovoltaic cell \cite{villalva2009comprehensive} as shown in Figure \ref{fig:pv2}. The latter is composed of a current source in parallel with a diode, a shunt resistance and a series resistance. 
\\
The DC current from the PV array (in Figure \ref{fig:pv2}) is calculated by the following equation \cite{villalva2009comprehensive}
\be \label{eq:I} I = I_{pv,cell} N_p - I_0 N_p [\exp(\frac{V + R_s I}{V_T a})-1]-\frac{V + R_s I}{R_p},\ee
where, $I$ is the PV-array output current, $V$ is the array output voltage, $I_{pv,cell}$ is the generated current from solar irradiance per cell, $I_0$ is the reverse saturation current, $V_T$ is the thermal voltage, $a$ is the diode  constant, $N_p$ is the number of PV-cells in parallel in the array, $R_p$ is the parallel resistance, and $R_s$ is the series resistance.\\[9mm] Furthermore
\begin{itemize}
    \item $V_T = \dfrac{N_s k T}{q}$,
    \item $I_{pv,cell} = (I_{pv,n} + K_I(T - T_n))\dfrac{G}{G_n}$,
    \item $I_0 = \dfrac{I_{sc,n} + K_I(T - T_n)}{\exp\left((V_{oc,n} + K_V(T - T_n))/aV_T \right) - 1}$
\end{itemize}
where $N_s$ is the number of PV-cells in series in the array, $k$ is the Boltzmann constant ($1.3806503 \cdot 10^{-23}$ [$J/K$]), $q$ is the electron charge ($1.60217646 \cdot 10^{-19} $ [$C$]), $K_I$, $K_V$ are the short-circuit current/temperature coefficient and the open circuit voltage/temperature coefficient, respectively, $T$ is the cell temperature in [$K$], $G$ is the solar irradiation in [$W/m^2$], $I_{pv,n}$, $I_{sc,n}$, $V_{oc}$, $T_n$, $G_n$ are the PV current, short circuit current, open circuit voltage, temperature, and solar irradiation at standard test conditions ($T_n = 298$ [$K$], $G_n = 1000$ [$W/m^2$]), respectively.
The cell temperature can be calculated from the air temperature with $ T_{cell} = T_{air} + \dfrac{NOCT - 20}{80} G$, where NOCT is the \textit{Nominal Operating Cell Temperature} (see \cite{temp}).\\[1mm]
We calculate the current from the above formula \cite{kees} and the parameters for a KC200GT Solar Array \cite{villalva2009comprehensive,kc200gt} (as in Figure \ref{fig:para}) with $N_s = 54$ and $N_p = 1$.
We can now compute the $I$-$V$ curve (e.g. Figure \ref{fig:I-Vcurve}) and determine the values for $I$ and $V$ where the power is maximized, the \textit{maximum power point} (MPP). 
The solar irradiation and temperature data are based on KNMI data \cite{knmi}. The data from the last three years (2013, 2014, and 2015) is used in order to calculate an average day over all days from all three years for the month June. Figure \ref{fig:ts} shows the solar irradiation and temperature for an average day in June based on the data from \cite{knmi}. 
\subsection*{Control of inverter power}  
Considering the simplified model, the DC current from the PV-arrays is converted to an AC signal given by
\begin{equation}
i_{pv}(t) = |i_{pv}| \cos(\omega t+\phi),
\end{equation}
where $\omega$ is the nominal grid frequency, $\phi$ is the angle phase and $|i_{pv}|= n I$ with $n$  the number of arrays at the inverter and $I$ is the output current of one array. We assume that $\phi$ (e.g. by a PI controller) and the current size $|i_{pv}|$ (e.g. with a proportional controller) are controlled in order to track the desired optimum power values $P^\ast$ and $Q^\ast$. Moreover, since IEEE 37 nodal voltages are in the MV (medium voltage) range and the PV panels are assumed to be at LV voltage level (220 V rms value), we assume an ideal transformer at each busbar such that $N=\frac{|v_{MV,grid}|}{|v_{LV,grid}|}$. Hence, based on the reference values $P^\ast$ and $Q^\ast$, we calculate
\begin{equation}
\begin{array}{rcl}
\phi_{L}^\ast&=&\arctan \frac{Q^\ast}{P^\ast},\\[1mm]
I_{L,MV}^\ast &=&\frac{P^\ast}{|v_{MV,grid}| \cos \arctan \frac{Q^\ast}{P^\ast}}.
\end{array}
\end{equation}
Considering the transformer between MV and LV, we obtain
$$I_L^\ast= N I_{L,MV}^\ast,$$ 
where $I_L$ is the size of the current $i_L(t)$ through the inductor $L$. For the simplified circuit of the inverter, we have 
\begin{equation}\label{eq:rl}
\begin{array}{rcl}
v_R &=& v_L + v_g,\\[1mm]
R (i_{pv}-i_{L})&=& L \frac{di_L}{dt} +v_g.
\end{array}
\end{equation}
Now, knowing $I_L^\ast$ and $\phi_{L}^\ast$, the variables $|i_{pv}|^\ast$, $\phi^\ast$ and their relevant control gain can be computed.
\section{Conclusions}\label{sec:con}
This document has presented the data for a single-phase distribution bus (based on IEEE 37 bus) together with the model and data for a PV inverter and active and reactive power loads.
\bibliographystyle{plain}
\bibliography{biblio}
\end{document}